\documentclass[twocolumn,prl]{revtex4}
\usepackage{graphics,graphicx}

\begin{document}
\title{Quantum continuum fluctuations in glassy perovskite Ca(Co$_{0.15}$Ru$_{0.85}$)O$_{3}$}
\author{Yiyao~Chen$^{1}$}
\author{A.~Dahal$^{1}$}
\author{J.~A.~Rodriguez-Rivera$^{2,3}$}
\author{Guangyong Xu$^{2}$}
\author{T. Heitmann$^{4}$}
\author{V.~Dugaev$^{5}$}
\author{A.~Ernst$^{6,7}$}
\author{D. J.~Singh$^{1}$}
\author{D. K.~Singh$^{1,*}$}
\affiliation{$^{1}$Department of Physics and Astronomy, University of Missouri, Columbia, MO, USA}
\affiliation{$^{2}$NIST Center for Neutron Research, Gaithersburg, MD, USA}
\affiliation{$^{3}$Department of Materials Science and Engineering, University of Maryland, College Park, MD,USA}
\affiliation{$^{4}$University of Missouri Research Reactor, University of Missouri, Columbia, MO, USA}
\affiliation{$^{5}$Department of Physics and Medical Engineering, Rzesz\'ow University of Technology, Rzesz\'ow, Poland}
\affiliation{$^{6}$Max-Planck-Institut f\"ur Mikrostrukturphysik, Weinberg 2, 06120 Halle, Germany}
\affiliation{$^{7}$Institut f\"ur Theoretische Physik, Johannes Kepler Universit\"at, 4040 Linz, Austria}
\affiliation{$^{*}$email: singhdk@missouri.edu}

\begin{abstract}

The quantum spin continuum and classical spin freezing, associated with a glassy state, represent two opposite extremes of a correlated electronic material. Here, we report the coexistence of a quantum spin continuum with a weak spin glass order in Co-doped CaRuO$_{3}$ perovskite near the chemical doping dependent metal-insulator transition boundary.
Inelastic neutron measurements on Ca(Co$_{0.15}$Ru$_{0.85}$)O$_{3}$
at low temperature, $T$ = 1.5 K, reveal a continuum spectrum in the $Q-E$ space due to uncorrelated spin fluctuations. This persists across the glass transition at $T_G \simeq$~23 K. Furthermore, scaling of the dynamic susceptibility yields a very small scaling coefficient $\alpha$  $\simeq$ 0.1, suggesting extreme locality of the dynamic properties. The experimental results indicate the realization of a narrow regime where the distinction between continuum dynamic behavior and glass-like regimes is reduced.

\end{abstract}

\maketitle

\section{I. Introduction}
The quantum continuum fluctuation, often found in spin liquid candidate materials, leads to a novel phase of matter in which contiguous non-correlated spin fluctuations construct a continuous spectrum in energy-momentum space. \cite{Balents,Coldea,Lee,Kivelson}
A purely classical spin glass state, characterized by freezing of dynamic activities of magnetic ions below the glass transition temperature is another extreme. \cite{Young}
While the continuum fluctuation implies a continuous distribution of spin relaxation rate, a spin glass exhibits a statistical distribution of relaxation times with a mean value corresponding to the glass transition temperature. \cite{Huser}
Relaxing this strict criteria may allow for a reversible cross-over
from one phase, spin glass, to another, dynamic phase of continuum fluctuation, as a function of temperature or other tuning parameter, such as magnetic field, or disorder.
Indeed, a quantum phase of spin glass was theorized to explain the
destruction of the Ising spin glass state in LiHo$_{0.167}$Y$_{0.833}$F$_{4}$ in a transverse field, \cite{Aeppli} causing the system to become quantum critical as $T \rightarrow$ 0 K. \cite{Sachdev1,Sachdev2}
Exploration of a new regime, manifesting the coexistence of both quantum and classical states, can lead to the elucidation of a new phase of matter. Here, we present a study in this direction.

\begin{figure}
\centering
\includegraphics[width=8.7 cm]{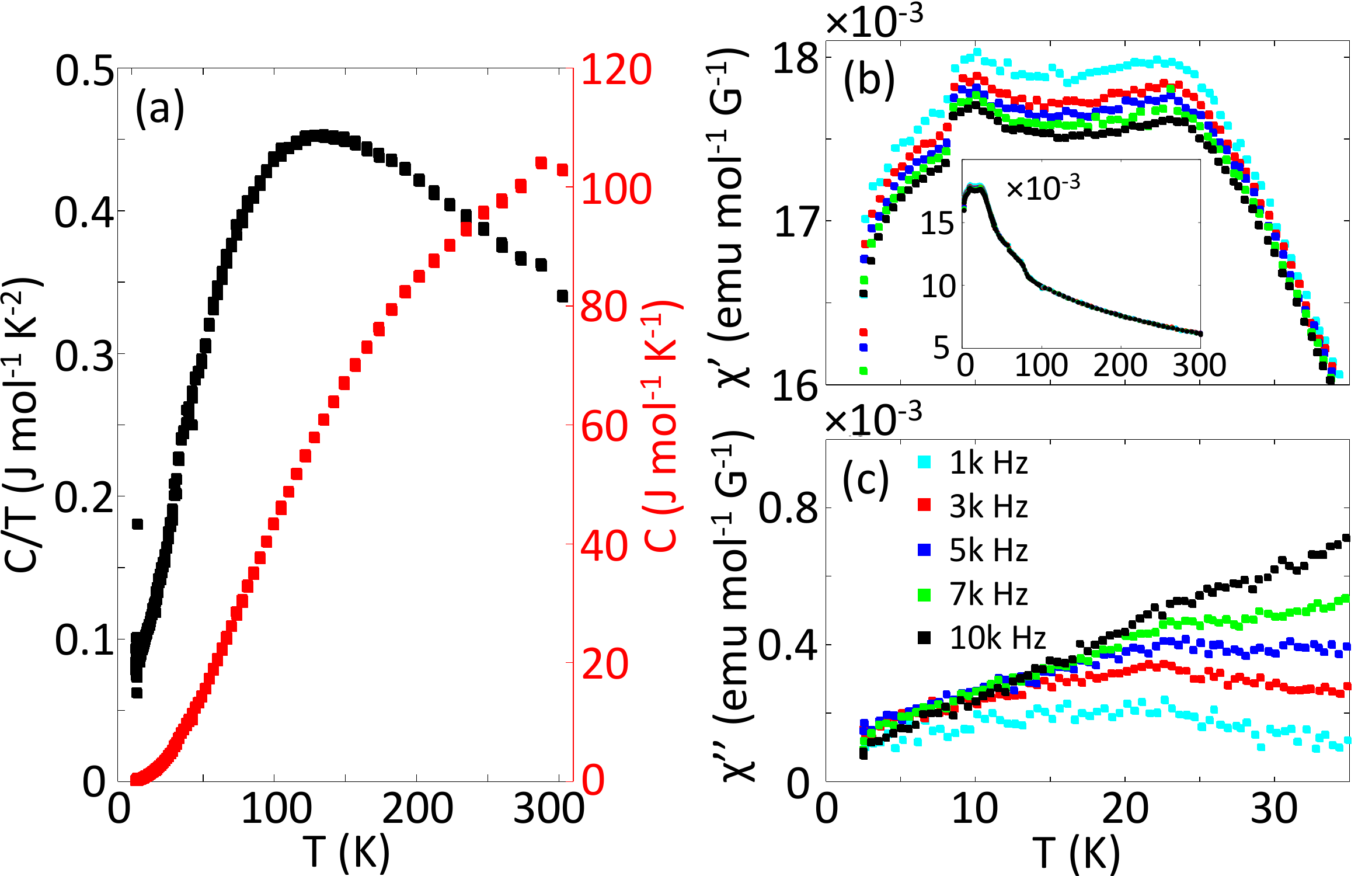} \vspace{-4mm}
\caption{Thermal and magnetic characterization of Ca(Co$_{0.15}$Ru$_{0.85}$)O$_{3}$. (a) Heat capacity data is shown here. No indication of any magnetic order is detected in this measurement. Rather, a broad peak, suggesting a band of relaxation times, is observed. (b) ac static susceptibility measurements exhibit a frequency-dependent cusp around $T \simeq$ 23 K, indicating the onset of a glassy phase in the system. (c) ac dynamic susceptibility measurements, exhibiting frequency dependent enhancement, complement the glassy characteristic observed in the static susceptibility data.
} \vspace{-4mm}
\end{figure}

\begin{figure*}
\centering
\includegraphics[width=17 cm]{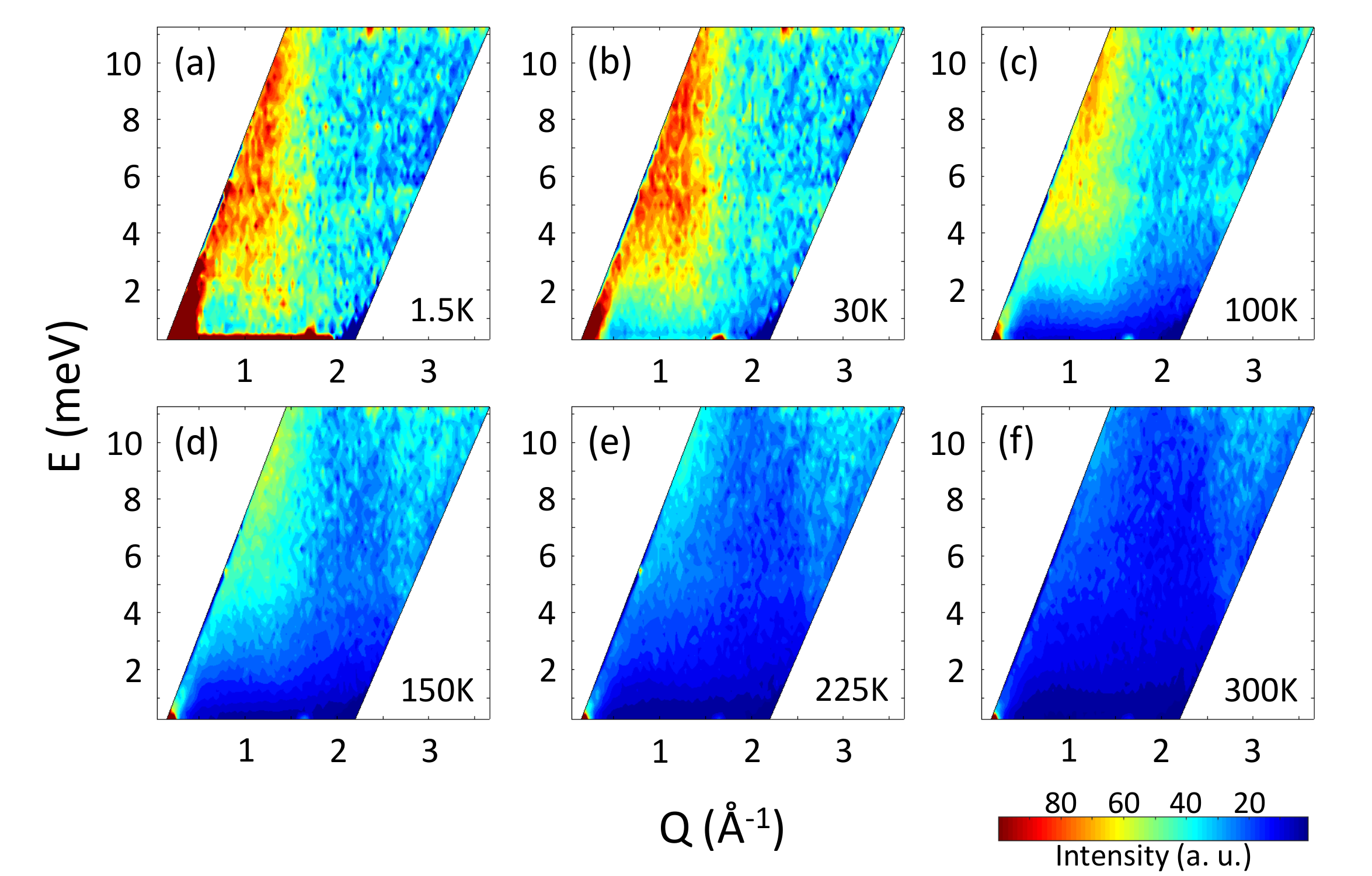} \vspace{-4mm}
\caption{Inelastic neutron scattering measurements of Ca(Co$_{0.15}$Ru$_{0.85}$)O$_{3}$. (a)-(f) $Q-E$ map contour plots, obtained on MACS cold spectrometer, at various temperatures. Inelastic measurements were performed with fixed final energy $E_f$ = 3.7 meV with energy resolution $\simeq$ 0.3 meV. The fluctuation spectrum becomes stronger as temperature is reduced,
thus manifesting the quantum characteristic of the dynamic properties.
Also, the absence of any spatial correlation in the dynamic structure factor makes it an uncorrelated continuum excitation. } \vspace{-4mm}
\end{figure*}

We report the coexistence of a weak spin glass phase with the continuum spin fluctuation, which becomes stronger as temperature decreases, in Co-doped CaRuO$_3$. The spin glass phase may be arising due to the disorder induced by the substitution of Ru by Co ions in the host compound. CaRuO$_{3}$ is a non-Fermi liquid metal, \cite{Gunasekera,Kikugawa,Mukuda}
apparently near a quantum critical point associated with the ferromagnetism
of Sr substituted material. \cite{mazin,He,khalifah}
Upon partial substitution of Ru by Co, it exhibits a metal-insulator transition at $\simeq$ 15\% Co (see Fig. S1 in the Supplementary Material\cite{supp}).
At x $>$ 0.15, Ca(Co$_{x}$Ru$_{1-x}$)O$_{3}$ exhibits insulating properties. However, Ca(Co$_{0.15}$Ru$_{0.85}$)O$_{3}$ (at x = 0.15) is not an insulator. Rather, it is on the verge of becoming an insulator.
In any case, the strong signatures of quantum criticality and the
suppression of ferromagnetism by quantum fluctuations in CaRuO$_3$ suggest
a competing order, perhaps an unidentified antiferromagnetic order,
\cite{shepard} or a more subtle quantum state, such as have been noted in models. \cite{Mizusaki,Meng,Senthil}
Therefore, Ca(Co$_{0.15}$Ru$_{0.85}$)O$_{3}$ is a prime candidate for 
investigation of the critical dynamic instability.

Using ac susceptibility, heat capacity and neutron scattering measurements on this system, we find the coexistence of a spin glass state, onset below $T_G \simeq$ 23 K, with a broad gapless quantum continuum at low temperature.
No trace of any type of magnetic order is detected in this material.
Even though the quantum spin fluctuations are $Q$-dependent, the scaling of dynamic susceptibility remains somewhat $Q$-independent for a large range of momentum vectors.
Most notably, the dynamic scaling coefficient, $\alpha$, is found to be very small, $\simeq$ 0.1, which suggests extreme locality of the dynamic behavior. \cite{Continentino}
The gapless continuum subsides significantly as temperature increases above $T\simeq$ 100 K.
Our observations are in direct contrast with the conventional understanding
that forbids a continuum-type dynamic behavior in the glassy phase of a magnetic material.

\section{II. Experimental}

The high purity polycrystalline samples of Ca(Co$_{0.15}$Ru$_{0.85}$)O$_{3}$
were synthesized by conventional solid-state reaction method in an oxygen-rich environment using ultra-pure ingredients of CoO, RuO$_{2}$ and CaCO$_{3}$. Starting materials were mixed in stoichiometric composition, pelletized and sintered at 950$^{o}$ for three days. The furnace-cooled samples were grinded, pelletized and sintered at 1000$^{o}$ for another three days. Samples were intentionally synthesized at slightly lower temperature and for longer duration to preserve the oxygen stoichiometry. The high quality of the sample is verified using powder X-ray diffraction, every single peak in the diffraction profile is identified with the orthorhombic CaRuO$_{3}$ structure (see Fig. S2 in the Supplementary Material\cite{supp}). 

Heat capacity and ac susceptibility measurements were performed using a Quantum Design Physical Properties Measurement System with a temperature range of 2-300 K.\cite{NIST} Inelastic measurements were performed on cold spectrometers multi-axis crystal spectrometer (MACS) and SPINS at the NIST Center for Neutron Research with fixed final neutron energies of 3.7 meV and 5 meV, respectively. At this final energy, the spectrometer's energy resolutions were determined to be $\simeq$ 0.3 meV and 0.28 meV, respectively. Elastic neutron scattering measurements were performed on SPINS and powder diffractometer PSD at the Missouri University Research Reactor.

\section{III. Results}

\subsection{A. Spin glass phase in Ca(Co$_{0.15}$Ru$_{0.85}$)O$_{3}$} 
Characterization of thermal and magnetic properties of Ca(Co$_{0.15}$Ru$_{0.85}$)O$_{3}$ was first performed using heat capacity, $C(T)$, and ac susceptibility measurements.
As shown in Fig. 1a, a very broad peak is observed in $C(T)/T$ vs. $T$.
The broad peak in the heat capacity plot may be arising due to the continuous relaxation of spin fluctuations.\cite{Singh}
The absence of any sharp increase in heat capacity or, Schottky-type enhancement at low temperature suggests the absence of any magnetic order or short-range ordered cluster formation in the system.\cite{Dood}
This is confirmed by the elastic neutron scattering measurements where no additional peak or diffuse scattering that might indicate the development of 
magnetic clusters is observed (see Fig. S3 in the Supplementary Material\cite{supp}).
Also, the elastic measurement data remains magnetic field independent up to $H$ = 10 T.
However, the static susceptibility data, shown in Fig. 1b, has a weak ac frequency dependent cusp around $T_G$ = 23 K.
Although magnetic susceptibility does not exhibit a significant change
between $T$ = 300 K to 2 K (see the inset in Fig. 1b), consistent with paramagnetism, the weak ac frequency dependent cusp hints of the glassiness in the system.
Similarly, the dynamic susceptibility data exhibits strong enhancement
at higher ac frequency near the transition,
as is often observed in spin glass type systems. \cite{Singh,Mydosh} While these results suggest glassy character due to the Co-substitution,
such a behavior is not very prominent. This could be occurring due to competing spin fluctuations.
Prior studies of Co-doped CaRuO$_{3}$ have also indicated the presence of a glassy phase in moderately and highly doped systems. \cite{He,Breard} There is a diversity in the observed nature of the glassy phase in Ca(Co$_{x}$Ru$_{1-x}$)O$_{3}$. It varies from a frequency-dependent sharp cusp in $\chi^{'}$ to a plateau-type region, following a cusp, below $T_G$.\cite{He,Breard} Nevertheless, the existence of weak spin glass phase is confirmed through multiple reports. Such peaks in $\chi^{'}$ have been also seen in Sr$_3$Ru$_2$O$_7$ where it is associated with the proximity to quantum criticality. \cite{grigera}

\subsection{B. Quantum continuum excitation in Ca(Co$_{0.15}$Ru$_{0.85}$)O$_{3}$} 
We performed detailed inelastic neutron scattering measurements on Ca(Co$_{0.15}$Ru$_{0.85}$)O$_{3}$ to investigate the spin dynamics at low temperature.
Figure 2 shows the $Q-E$ maps at various temperatures obtained on the
MACS spectrometer.
Experimental data are background corrected and thermally balanced using the equation $Intensity(\chi^"(Q,E,T))=S(Q,E,T)\pi(1-e^{-E/k_BT})$.
At $T$ = 1.5 K, the entire energy-momentum space, under experimental investigation, is occupied by dynamic spectrum of the system.

Ca(Co$_{0.15}$Ru$_{0.85}$)O$_{3}$ shows gapless continuous spin fluctuations
to at least $E$ = 11.25 meV (maximum accessible energy) at sufficiently low temperature (Fig. 2a-c).
Moreover, the continuum spectrum becomes stronger as temperature decreases,
consistent with a quantum phenomenon.
However, these gapless fluctuations exhibit subtle $Q$-dependence.
As temperature increases above $T \simeq$ 150 K (see Fig. 2d),
the continuous spectrum starts developing into a gapped excitation of collective nature.
The collective excitation becomes weaker as temperature increases further.
The fluctuation subsides significantly at $T \geq$ 200 K but seems to maintain weak short-range dynamic correlations, centered around $Q$ = 1 and 3 \AA$^{-1}$ (Fig. 2e-f).
The distinct temperature dependencies of the dynamic properties signifies it as a magnetic phenomenon and the high temperature behavior is indicative of antiferromagnetic fluctuations.

\begin{figure}
\centering
\includegraphics[width=8.9 cm]{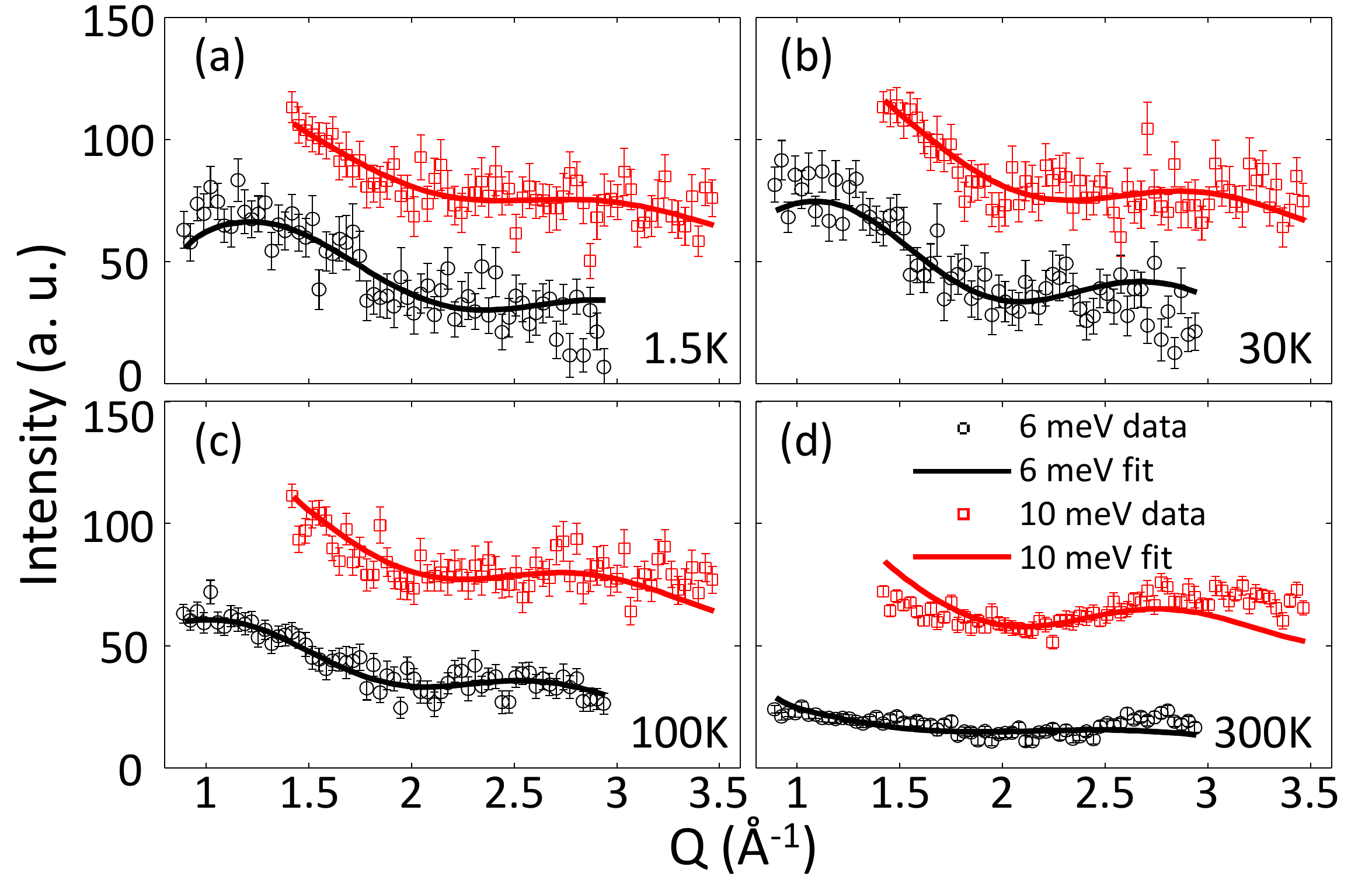} \vspace{-4mm}
\caption{Quantitative estimation of spatial correlation at different
excitation energies. (a-b) To obtain quantitative information about the correlation length of dynamic properties, we plot the cuts across fixed energy of $E$ = 6 and 10 meV as a function of Q and fit them using equation (1). Fitting at low temperature suggests that the correlation does not extend to the nearest neighbor lattice site. Experimental data at low and high energy are separated by a finite amount for clarity. (c-d) As temperature increases further, the spatial correlation extends to the nearest neighbor site of the lattice. At $T$ = 300 K, experimental data becomes indistinguishable from the background at low energy. However, a weak but finite nearest neighbor correlation is detected at higher energy. Error bar represents one standard deviation in the data.
} \vspace{-4mm}
\end{figure}

As mentioned previously, stoichiometric CaRuO$_{3}$ apparently manifests a near quantum critical state as $T \rightarrow$ 0 K. \cite{Gunasekera}
It is also evidenced by the non-Fermi liquid scaling over a sizable temperature range.\cite{cao}
Upon partial substitution of Ru by Co-ions, the spin dynamics that were
originally confined to the quasi-elastic excitations demonstrate a collective behavior as x $\rightarrow$ 0.12.
Clearly, the Co doping causes a dramatic effect on the spin dynamics in
Ca(Co$_{x}$Ru$_{1-x}$)O$_{3}$.
At the same time, it also introduces disorder into the system.
Generally, two types of excitations are observed in a disordered system:
(a) spin excitations between equivalent degenerate configurations
via energy barrier crossing, and (b) magnon excitation within the equilibrium configuration.
\cite{Rhyne} The latter possibility can only be realized if the system exhibits a continuous symmetry, as found in the Heisenberg system.
Since a magnon excitation is well-defined in the momentum space,
it does not explain the continuum spin dynamics in Ca(Co$_{0.15}$Ru$_{0.85}$)O$_{3}$, as shown in Fig. 2a.
The first explanation then seems more plausible.
To quantify the size of the dynamic spin correlation and understand the scope of activation across the energy barriers, we fit the two-dimensional cuts along momentum at fixed energies of $E$ = 6 meV and 10 meV at various temperatures between $T$ = 1.5 K and 300 K using the following expression,\cite{Furrer}
\begin{eqnarray}
{I (q)}&{\propto}& {f(q)}^{2}.{\sum_{ij}} [1 + {(-1)^ {\Delta {S_T}}} cos({\textbf{q}.\textbf{r}_{ij}})]. {e^{-{E_a / k_B T}}}
\end{eqnarray},
where $f(q)$ is the magnetic form factor of cobalt, $r_{ij}$ = $r_i$ - $r_j$
(distance between two neighboring ions forming a pair), $S_T$ = $S_i + S_j$ and $E_a$ is the activation energy. The sum runs over all possible pairs.
The momentum vector is powder averaged for the fitting purposes. Under the selection rule, allowed values of $\Delta${$S_T$} are 0 and $\pm$1.
As described in Ref.(\cite{Furrer}), the oscillatory nature of the expression yields information about the distance between the correlated spins.
As shown in Fig. 3, the fitted value of $r_{ij}$ at $T$ = 1.5 K varies between 3.4 \AA{} to 3.6 \AA{} at different energies. The estimated value of $r_{ij}$ is less than the nearest inter-atomic distance or the lattice parameters of Ca(Co$_{0.15}$Ru$_{0.85}$)O$_{3}$. Therefore, the system does not develop any spatial correlation, thus hints at uncorrelated spin fluctuation of Co-ions in the system.
Application of this analysis to the momentum dependence of dynamic properties at higher temperatures, $T$ = 50 K - 300 K, yields $r_{ij}$ = \{3.6, 4.1\} \AA.
This is comparable to the nearest inter-atomic distance of 3.86 $\AA$.
Thus, the system tends to develop short-range dynamic spatial correlation with nearest neighbor at higher temperature. Given that Ca(Co$_{0.15}$Ru$_{0.85}$)O$_{3}$ exhibits a weak glassy phase below $T_G$ = 23 K, this observation indicates the coexistence of a quantum continuum fluctuation and a spin glass phase at low temperature.

\begin{figure}
\centering
\includegraphics[width=8.8 cm]{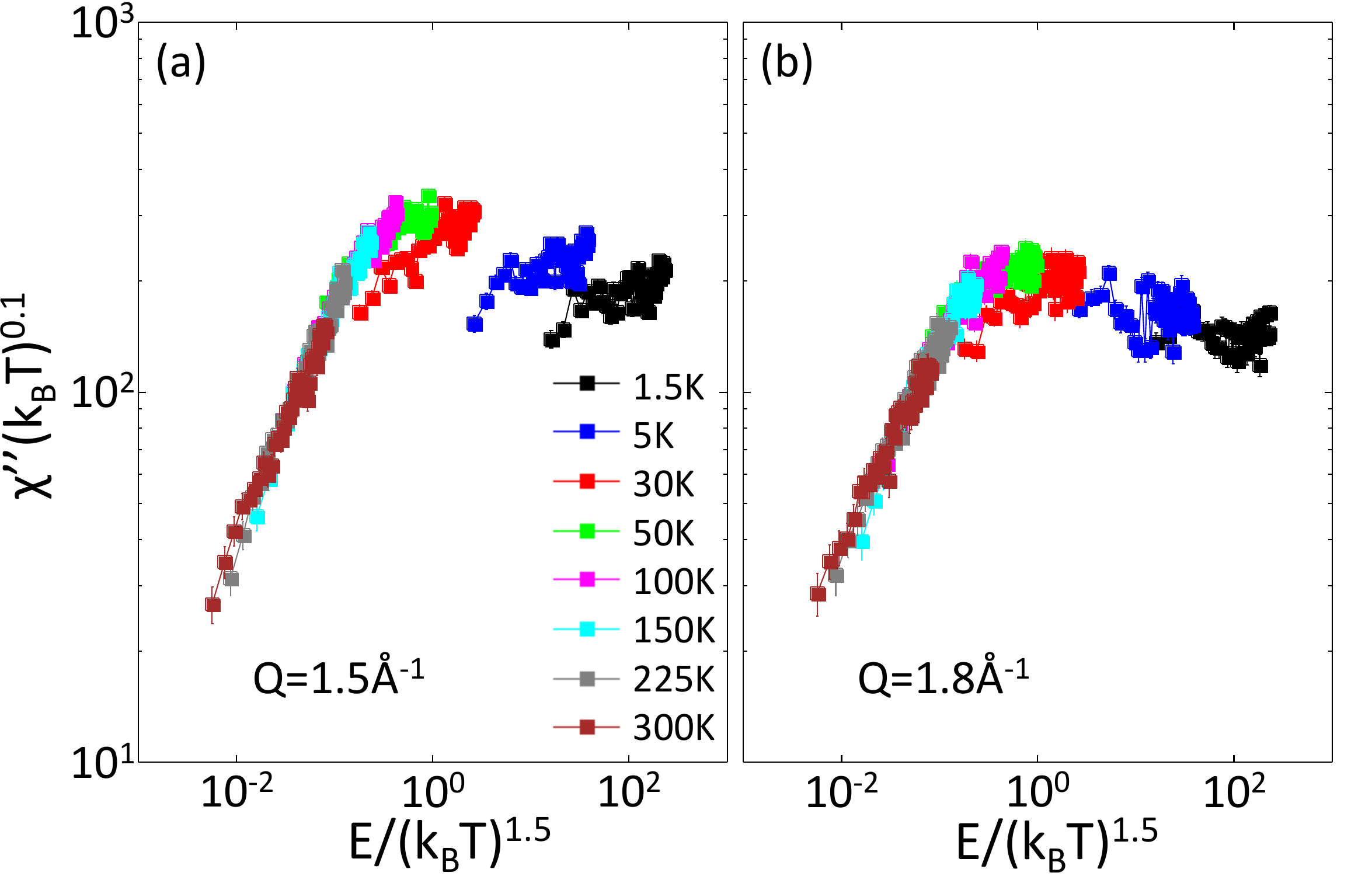} \vspace{-4mm}
\caption{Scaling of dynamic susceptibilities. (a) At $Q$ = 1.5 \AA$^{-1}$ and (b) $Q$ = 1.8 \AA$^{-1}$. Two behaviors are observed: very small scaling coefficient, $\alpha \simeq$~0.1, suggesting extreme locality of the dynamic properties, and near Q-independent scaling of dynamic susceptibilities
despite enhanced fluctuations at low Q. Both properties hint of uncorrelated local fluctuations in the system.
} \vspace{-4mm}
\end{figure}

\subsection{C. Extreme locality of dynamic properties} 
Spin fluctuations below a glass transition temperature are not uncommon.
There are several examples, especially in the geometrically frustrated pyrochlore materials.\cite{Dunsiger}
Although Ca(Co$_{0.15}$Ru$_{0.85}$)O$_{3}$ only exhibits a weak glassy
character, the nature of liquid-like uncorrelated spin fluctuations is highly unexpected.
It lends support to the view that the underlying spin dynamics, which persist to very high energy (corresponding to very short relaxation time),
competes against the static order.
Further information can be obtained about this competing behavior by
scaling analysis of the dynamic parameter, namely the dynamic susceptibility
($\chi^"$).\cite{Continentino} In Fig. 4, we plot $\chi^"$.($k_B$T)$^{\alpha}$ vs. ($E/k_B T$)$^{\beta}$ for two different $Q$-values.\cite{Singh2}
Best correlation between dynamic susceptibility data is obtained for scaling coefficients $\alpha$ $\simeq$ 0.1 and $\beta$  = 1.5 (see Fig. S4 in the Supplementary Material\cite{supp} for the scaling analysis for different values of $\alpha$ and $\beta$).
Two important findings are obtained from this exercise: First, the dynamic susceptibilities indeed exhibit scaling behavior. However, the scaling coefficient $\alpha$ is very small, 0.1. Second, the scaling of $\chi^"$ is somewhat $Q$-independent for these coefficients (also see Fig. S5 in the Supplementary Material\cite{supp}). While more detailed analysis is clearly required to fully understand the origin of such a small value of $\alpha$, these analyses corroborate the extremely local nature of the spin dynamics.\cite{Continentino} Perhaps, it explains why the system does not develop the finite size clusters that are necessary for the expansion of the nascent spin glass phase. 

\section{IV. Discussion}

Spin glass and quantum continuum fluctuations are non-congruent properties of matter. One may anticipate crossing over from a spin glass phase to a continuum fluctuation phase by reducing the hydrodynamic viscosity (here the damping of spin fluctuations) of the material, thus melting the underlying short-range static order. Even though Ca(Co$_{0.15}$Ru$_{0.85}$)O$_{3}$ shows the quantum spin continuum at low temperature, it strives to develop a glassy phase. The strong spin fluctuations compete against this tendency of the system. It is helpful to consider the spin dynamics that would be due to cluster formation in the system. Generally, the spin dynamics involves a transition between various energy levels, given by $E (S_T)$ = $J S_T(S_T + 1)$ in the spin clusters.\cite{Furrer2,Rhyne} As temperature increases, higher energy levels become populated. As a result, new excitations appear in the inelastic spectrum. In this case, neither a well-defined excitation is detected at any temperature nor does any new inelastic peak emerge. Rather, the dynamic properties become weaker as temperature increases,
which is reminiscent of the continuum spin fluctuations that develop
into a quantum phenomenon as $T \rightarrow$ 0 K. At higher temperature, the moment tends to develop nearest neighbor correlation, as evidenced from the quantitative analysis in Fig. 3. Thus, the extremely local nature of moment fluctuation, which makes it a continuum phenomenon, breaks down. The system starts populating higher energy states, given by the total angular momentum, at higher temperature and the excitation becomes gapped. The observed dynamic behavior in Ca(Co$_{0.15}$Ru$_{0.85}$)O$_{3}$ is significantly different from the quasi-critical behavior in CaRuO$_{3}$ where the width of inelastic peak approaches the instrument resolution as T $\rightarrow$ 0 K.\cite{Gunasekera} Also, no continuum fluctuation is observed in the stoichiometric composition CaRuO$_{3}$ (see Fig. S6 in the Supplementary Material\cite{supp}). 

A glassy system is characterized by its viscosity. Substitution of ruthenium by cobalt in Ca(Co$_{x}$Ru$_{1-x}$)O$_{3}$ tends to cause a spin glass behavior at x $\geq$0.1 in this system.\cite{He,Breard} Unlike a conventional spin glass system where the spin freezing is well pronounced below the glass transition temperature,\cite{Young} Ca(Co$_{x}$Ru$_{1-x}$)O$_{3}$ manifests weak glassy character.
We argue that the strong spin fluctuations at low temperature significantly
weakens the relaxation time of the nascent glassy phase and prohibits it from developing a static order. The complementary calculation of change in density of states suggests that Co doping strongly affects the CaRuO$_3$ host, both through changed electron count and the combination of Co moments and strong hybridization of Co and neighboring Ru $d$ states (see Fig. S7 and associated text in the Supplementary Material\cite{supp}). Therefore, it is remarkable that alloying with Co, which produces strong scattering, leads to enhanced local magnetic fluctuations.  Our work provides new information on the development of quantum mechanical properties, as manifested by the quantum continuum fluctuation, below the spin glass transition in a system. Future research works on the investigation of possible correlation between quantum continuum and a quantum phase of spin glass, as proposed by Sachdev et al. many years ago, are highly desirable.

\section{Acknowledgements}
The research at MU was supported by the Department of Energy, Office of Science, Office of Basic Energy Sciences under the grant no. DE-SC0014461. This work utilized neutron scattering facilities, supported in part by the Department of Commerce.

\clearpage


\begin{thebibliography}{99}
\bibitem{NIST} Department of Commerce and National Institute of Standards and Technology do not recommend any commercial product.
\end{thebibliography}
\end{document}